\documentclass[conference,a4paper]{IEEEtran}
\usepackage[left=0.7in, right=0.7in, bottom=1in, top=0.75in]{geometry}

\usepackage{multicol}
\usepackage{graphicx}
\graphicspath{{Figures/}}
\usepackage{epstopdf}
\usepackage{acronym}
\usepackage{amsmath}
\usepackage{color, colortbl}
\usepackage{subfigure}
\usepackage[english]{babel}
\usepackage{blindtext}
\usepackage{algorithm} 
\usepackage{algorithmic} 
\usepackage{multirow}
\usepackage{color}
\usepackage{subfigure}
\usepackage{balance}
\usepackage{caption}
\usepackage{amsfonts}

\begin{document}

\newacro{3GPP}{third generation partnership project}
\newacro{4G}{4{th} generation}
\newacro{5G}{5{th} generation}

\newacro{ADC}{analogue-to-digital conversion}
\newacro{AED}{accumulated euclidean distance}
\newacro{AI}{artificial intelligence}
\newacro{ASE}{amplified spontaneous emission}
\newacro{ASIC}{application specific integrated circuit}
\newacro{AWG}{arbitrary waveform generator}
\newacro{AWGN}{additive white Gaussian noise}
\newacro{A/D}{analog-to-digital}

\newacro{B2B}{back-to-back}
\newacro{BCF}{bandwidth compression factor}
\newacro{BCJR}{Bahl-Cocke-Jelinek-Raviv}
\newacro{BDM}{bit division multiplexing}
\newacro{BED}{block efficient detector}
\newacro{BER}{bit error rate}
\newacro{Block-SEFDM}{block-spectrally efficient frequency division multiplexing}
\newacro{BLER}{block error rate}
\newacro{BPSK}{binary phase shift keying}
\newacro{BS}{base station}
\newacro{BSS}{best solution selector}
\newacro{BU}{butterfly unit}

\newacro{CapEx}{capital expenditure}
\newacro{CA}{carrier aggregation}
\newacro{CBS}{central base station}
\newacro{CC}{component carriers}
\newacro{CCDF}{complementary cumulative distribution function}
\newacro{CCs}{component carriers}
\newacro{CD}{chromatic dispersion}
\newacro{CDF}{cumulative distribution function}
\newacro{CDI}{channel distortion information}
\newacro{CDMA}{code division multiple access}
\newacro{CI}{constructive interference}
\newacro{CIR}{carrier-to-interference power ratio}
\newacro{CMOS}{complementary metal-oxide-semiconductor}
\newacro{CNN}{convolutional neural network}
\newacro{CoMP}{coordinated multiple point}
\newacro{CO-SEFDM}{coherent optical-SEFDM}
\newacro{CP}{cyclic prefix}
\newacro{CPE}{common phase error}
\newacro{CRVD}{conventional real valued decomposition}
\newacro{CR}{cognitive radio}
\newacro{CRC}{cyclic redundancy check}
\newacro{CS}{central station}
\newacro{CSI}{channel state information}
\newacro{CSPR}{carrier to signal power ratio}
\newacro{CWT}{continuous wavelet transform}
\newacro{C-RAN}{cloud-radio access networks}

\newacro{DAC}{digital-to-analogue conversion}
\newacro{DBP}{digital backward propagation}
\newacro{DC}{direct current}
\newacro{DCT}{discrete cosine transform}
\newacro{DDC}{digital down-conversion}
\newacro{DDO-OFDM}{directed detection optical-OFDM}
\newacro{DDO-OFDM}{direct detection optical-OFDM}
\newacro{DDO-SEFDM}{directed detection optical-SEFDM}
\newacro{DFB}{distributed feedback}
\newacro{DFDMA}{distributed FDMA}
\newacro{DFT}{discrete Fourier transform}
\newacro{DFrFT}{discrete fractional Fourier transform}
\newacro{DMA}{direct memory access}
\newacro{DMRS}{demodulation reference signal}
\newacro{DOFDM}{dense orthogonal frequency division multiplexing}
\newacro{DP}{dual polarization}
\newacro{DPC}{dirty paper coding}
\newacro{DSB}{double sideband}
\newacro{DSL}{digital subscriber line}
\newacro{DSP}{digital signal processors}
\newacro{DVB}{digital video broadcast}
\newacro{DWT}{discrete wavelet transform}
\newacro{D/A}{digital-to-analog}

\newacro{ECC}{error correcting codes}
\newacro{ECL}{external-cavity laser}
\newacro{EDFA}{erbium doped fiber amplifier}
\newacro{EE}{energy efficiency}
\newacro{eMBB}{enhanced mobile broadband}
\newacro{eNB-IoT}{enhanced NB-IoT}
\newacro{EPA}{extended pedestrian A}
\newacro{EVM}{error vector magnitude}

\newacro{Fast-OFDM}{fast-orthogonal frequency division multiplexing}
\newacro{FBMC}{filterbank based multicarrier }
\newacro{FCE}{full channel estimation}
\newacro{FD}{fixed detector}
\newacro{FDD}{frequency division duplexing}
\newacro{FDM}{frequency division multiplexing}
\newacro{FDMA}{frequency division multiple access}
\newacro{FE}{full expansion}
\newacro{FEC}{forward error correction}
\newacro{FEXT}{far-end crosstalk}
\newacro{FF}{flip-flop}
\newacro{FFT}{fast Fourier transform}
\newacro{FIFO}{first in first out}
\newacro{F-OFDM}{filtered-orthogonal frequency division multiplexing}
\newacro{FPGA}{field programmable gate array}
\newacro{FrFT}{fractional Fourier transform}
\newacro{FSD}{fixed sphere decoding}
\newacro{FSD-MNSF}{FSD-modified-non-sort-free}
\newacro{FSD-NSF}{FSD-non-sort-free}
\newacro{FSD-SF}{FSD-sort-free}
\newacro{FSK}{frequency shift keying}
\newacro{FTN}{faster than Nyquist}
\newacro{FTTB}{fiber to the building}
\newacro{FTTC}{fiber to the cabinet}
\newacro{FTTdp}{fiber to the distribution point}
\newacro{FTTH}{fiber to the home}

\newacro{GB}{guard band}
\newacro{GFDM}{generalized frequency division multiplexing}
\newacro{GPU}{graphics processing unit}
\newacro{GSM}{global system for mobile communication}
\newacro{GUI}{graphical user interface}

\newacro{HC-MCM}{high compaction multi-carrier communication}
\newacro{HPA}{high power amplifier}

\newacro{IC}{integrated circuit}
\newacro{ICI}{inter carrier interference}
\newacro{ID}{iterative detection}
\newacro{IDCT}{inverse discrete cosine transform}
\newacro{IDFT}{inverse discrete Fourier transform}
\newacro{IDFrFT}{inverse discrete fractional Fourier transform}
\newacro{ID-FSD}{iterative detection-FSD}
\newacro{ID-SD}{ID-sphere decoding}
\newacro{IF}{intermediate frequency}
\newacro{IFFT}{inverse fast Fourier transform}
\newacro{IFrFT}{inverse fractional Fourier transform}
\newacro{IMD}{intermodulation distortion}
\newacro{IoT}{internet of things}
\newacro{IOTA}{isotropic orthogonal transform algorithm}
\newacro{IP}{intellectual property}
\newacro{ISC}{interference self cancellation}
\newacro{ISI}{inter symbol interference}

\newacro{LDPC}{low density parity check}
\newacro{LFDMA}{localized FDMA}
\newacro{LLR}{log-likelihood ratio}
\newacro{LNA}{low noise amplifier}
\newacro{LO}{local oscillator}
\newacro{LOS}{line-of-sight}
\newacro{LPWAN}{low power wide area network}
\newacro{LS}{least square}
\newacro{LTE}{long term evolution}
\newacro{LTE-Advanced}{long term evolution-advanced}
\newacro{LUT}{look-up table}

\newacro{MA}{multiple access}
\newacro{MAC}{media access control}
\newacro{MASK}{m-ary amplitude shift keying}
\newacro{MCM}{multi-carrier modulation}
\newacro{MC-CDMA}{multi-carrier code division multiple access}
\newacro{MCS}{modulation and coding scheme}
\newacro{MF}{matched filter}
\newacro{MIMO}{multiple input multiple output}
\newacro{ML}{maximum likelihood}
\newacro{MLSD}{maximum likelihood sequence detection}
\newacro{MMF}{multi-mode fiber}
\newacro{MMSE}{minimum mean squared error}
\newacro{mMTC}{massive machine-type communication}
\newacro{MNSF}{modified-non-sort-free}
\newacro{MOFDM}{masked-OFDM}
\newacro{MRVD}{modified real valued decomposition}
\newacro{MS}{mobile station}
\newacro{MSE}{mean squared error}
\newacro{MTC}{machine-type communication}
\newacro{MUSA}{multi-user shared access}
\newacro{MU-MIMO}{multi-user multiple-input multiple-output}
\newacro{MZM}{Mach-Zehnder modulator}
\newacro{M2M}{machine to machine}

\newacro{NB-IoT}{narrowband IoT}
\newacro{NEXT}{near-end crosstalk}
\newacro{NG-IoT}{next generation IoT}
\newacro{NLOS}{non-line-of-sight}
\newacro{NN}{neural network}
\newacro{NOFDM}{non-orthogonal frequency division multiplexing}
\newacro{NOMA}{non-orthogonal multiple access}
\newacro{NoFDMA}{non-orthogonal frequency division multiple access}
\newacro{NP}{non-polynomial}
\newacro{NR}{new radio}
\newacro{NSF}{non-sort-free}
\newacro{NWDM}{Nyquist wavelength division multiplexing }
\newacro{Nyquist-SEFDM}{Nyquist-spectrally efficient frequency division multiplexing}

\newacro{OBM-OFDM}{orthogonal band multiplexed OFDM}
\newacro{OF}{optical filter}
\newacro{OFDM}{orthogonal frequency division multiplexing}
\newacro{OFDMA}{orthogonal frequency division multiple access}
\newacro{OMA}{orthogonal multiple access}
\newacro{OpEx}{operating expenditure}
\newacro{OQAM}{offset-QAM}
\newacro{OSI}{open systems interconnection}
\newacro{OSNR}{optical signal-to-noise ratio}
\newacro{OSSB}{optical single sideband}
\newacro{OTA}{over-the-air}
\newacro{Ov-FDM}{Overlapped FDM}
\newacro{O-SEFDM}{optical-spectrally efficient frequency division multiplexing}
\newacro{O-FOFDM}{optical-fast orthogonal frequency division multiplexing}
\newacro{O-OFDM}{optical-orthogonal frequency division multiplexing}

\newacro{PA}{power amplifier}
\newacro{PAPR}{peak-to-average power ratio}
\newacro{PCE}{partial channel estimation}
\newacro{PD}{photodiode}
\newacro{PDF}{probability density function}
\newacro{PDP}{power delay profile}
\newacro{PDMA}{polarisation division multiple access}
\newacro{PDM-OFDM}{polarization-division multiplexing-OFDM}
\newacro{PDM-SEFDM}{polarization-division multiplexing-SEFDM}
\newacro{PDSCH}{physical downlink shared channel}
\newacro{PE}{processing element}
\newacro{PED}{partial Euclidean distance}
\newacro{PMD}{polarization mode dispersion}
\newacro{PON}{passive optical network}
\newacro{PPM}{parts per million}
\newacro{PRB}{physical resource block}
\newacro{PSD}{power spectral density}
\newacro{PU}{primary user}
\newacro{PXI}{PCI extensions for instrumentation}
\newacro{P/S}{parallel-to-serial}

\newacro{QAM}{quadrature amplitude modulation}
\newacro{QoS}{quality of service}
\newacro{QPSK}{quadrature phase-shift keying}

\newacro{RAUs}{remote antenna units}
\newacro{RBW}{resolution bandwidth}
\newacro{RF}{radio frequency}
\newacro{RMS}{root mean square}
\newacro{RoF}{radio-over-fiber}
\newacro{ROM}{read only memory}
\newacro{RRC}{root raised cosine}
\newacro{RSC}{recursive systematic convolutional}
\newacro{RTL}{register transfer level}
\newacro{RVD}{real valued decomposition}

\newacro{ScIR}{sub-carrier to interference ratio}
\newacro{SCMA}{sparse code multiple access}
\newacro{SC-FDMA}{single carrier frequency division multiple access}
\newacro{SC-SEFDMA}{single carrier spectrally efficient frequency division multiple access}
\newacro{SD}{sphere decoding}
\newacro{SDP}{semidefinite programming}
\newacro{SDR}{software defined radio}
\newacro{SE}{spectral efficiency}
\newacro{SEFDM}{spectrally efficient frequency division multiplexing}
\newacro{SEFDMA}{spectrally efficient frequency division multiple access} 
\newacro{SF}{sort-free}
\newacro{SIC}{successive interference cancellation}
\newacro{SiGe}{silicon-germanium}
\newacro{SINR}{signal-to-interference-plus-noise ratio}
\newacro{SISO}{single-input single-output}
\newacro{SMF}{single mode fiber}
\newacro{SNR}{signal-to-noise ratio}
\newacro{SP}{shortest-path}
\newacro{SRS}{sounding reference signal}
\newacro{SSB}{single-sideband}
\newacro{SSBI}{signal-signal beat interference}
\newacro{SSMF}{standard single mode fiber}
\newacro{STBC}{space time block coding}
\newacro{STO}{symbol timing offset}
\newacro{SU}{secondary user}
\newacro{SVD}{singular value decomposition}
\newacro{SVR}{singular value reconstruction}
\newacro{S/P}{serial-to-parallel}

\newacro{TDD}{time division duplexing}
\newacro{TDMA}{time division multiple access }
\newacro{TFP}{time frequency packing}
\newacro{THP}{Tomlinson-Harashima precoding}
\newacro{TOFDM}{truncated OFDM}
\newacro{TSVD}{truncated singular value decomposition}
\newacro{TSVD-FSD}{truncated singular value decomposition-fixed sphere decoding}

\newacro{UCR}{user compression ratio}
\newacro{UFMC}{universal-filtered multi-carrier}
\newacro{URLLC}{ultra-reliable and low-latency communication}
\newacro{USRP}{universal software radio peripheral}

\newacro{VDSL}{very-high-bit-rate digital subscriber line}
\newacro{VDSL2}{very-high-bit-rate digital subscriber line 2}
\newacro{VHDL}{very high speed integrated circuit hardware description language}
\newacro{VLC}{visible light communication}
\newacro{VLSI}{very large scale integration}
\newacro{VOA}{variable optical attenuator}
\newacro{VP}{vector perturbation}
\newacro{VSSB-OFDM}{virtual single-sideband OFDM}

\newacro{WAN}{wide area network}
\newacro{WCDMA}{wideband code division multiple access}
\newacro{WDM}{wavelength division multiplexing}
\newacro{WiFi}{wireless fidelity}
\newacro{WiGig}{Wireless Gigabit Alliance}
\newacro{WiMAX}{Worldwide interoperability for Microwave Access}
\newacro{WSS}{wavelength selective switch}

\newacro{ZF}{zero forcing}
\newacro{ZP}{zero padding}


\title{Non-Orthogonal Waveforms in Secure Communications}
\author{\IEEEauthorblockN{ Tongyang Xu}
 \IEEEauthorblockA{Department of Electronic and Electrical Engineering,
University College London, London, UK\\
 Email: {tongyang.xu.11@ucl.ac.uk}}}

\maketitle

\begin{abstract}

This work investigates the possibility of using non-orthogonal multi-carrier waveforms to defend against eavesdropping attacks. The sophisticated detection required for non-orthogonal signals provides a natural defence mechanism in secure communications. However, brute-force tactics such as maximum likelihood detection would break the defence by attempting all possible solutions. Thus, a waveform scaling strategy is proposed to scale up the number of non-orthogonally packed sub-carriers, which complicates signal detections and prevents eavesdropping. In addition, a waveform tuning strategy is proposed to intentionally tune waveform parameters to enhance feature similarity. Therefore, eavesdroppers would be confused to misidentify signals resulting in subsequent detection failures.

\end{abstract}


\section{Introduction} \label{sec:introduction}

The open nature of wireless environment makes radio communications vulnerable \cite{adversarial_survey_2016} to eavesdropping data interception \cite{eavesdropping_CM_2015}. Defence strategies \cite{adversarial_JASC_2018}, such as millimetre wave, beamforming, artificial noise, security coding and directional modulation are proposed to mitigate the eavesdropping. Existing defence solutions are more likely dependent on surrounding channel environment and therefore are not robust in time-variant multipath fading channels when \ac{CSI} is imperfectly known \cite{security_CM_CSI_2015}. Traditional theoretical research prefers assuming perfect CSI or some other ideal assumptions, which makes theoretically achieved discoveries unrealistic in practical field experiment tests. Secure multiple user access is hardly implementable when legitimate users and eavesdroppers are close in space \cite{eavesdropping_CM_2015}, which is limited by imperfect beamforming leakages. In addition, the typical \ac{NOMA} based solution \cite{NOMA_TCom_2017} has risks of information leakages since one user is allowed to decode signals from other users. Traditional ways to extend secure communication coverage would rely on error correction coding \cite{security_TIT_2007} while its power and throughput efficiency is limited. Artificial noise enabled security is treated as an efficient defence solution \cite{security_AN_TWC_2008}. However, extra power would be wasted to generate noise and security reliability is compromised. Data encryption \cite{Zhang2017DesignOA}, widely used at link or transport layers, is also applicable to enhance physical layer security. However, its applications are limited and unrealistic in low-cost consumer-level products. Moreover, encrypted data could be captured by eavesdroppers and processed offline using brute-force tactics. Therefore, advanced defence countermeasures are needed to replace or complement traditional channel dependent physical layer security solutions. 

With the development of artificial intelligence, machine learning/deep learning based adversarial attacks \cite{adversarial_attack_ICC_2018, adversarial_attack_CL_2019} become more destructive than typical eavesdropping attacks. As defined in \cite{adversarial_attack_CL_2019}, adversarial attacks are divided into white-box attack and black-box attack. The white-box attack indicates that the adversary has perfect knowledge of the signal formats while the black-box attack assumes no knowledge about the signal formats. Practically, the signal format knowledge is not known by an adversary. Therefore, learning signal features will be the first step in the black-box attack. Work in \cite{adversarial_sagduyu_2019} explains three main types of attack termed inference attack, evasion attack and causative attack. A defence strategy is proposed in \cite{adversarial_attack_ICC_2018} where a legitimate user can use fake labels to fool an adversary attacker. In this case, the attacker cannot intelligently train a reliable signal classifier at the inference attack stage. This is equivalent to a causative attack from a legitimate user to the attacker by falsifying the attacker's training data. However, the throughput would be reduced because of the fake labels transmission. Therefore, maintaining a balanced throughput and security quality is a challenge to be solved. 

A non-orthogonal waveform \ac{SEFDM}, unlikely to be identified by eavesdroppers, is crafted for enhancing physical layer security. The research of the non-orthogonal waveform is traced back to 2003 \cite{SEFDM2003}. Unlike the multicarrier \ac{OFDM} signal, SEFDM packs sub-carriers closer by violating the orthogonality leading to either bandwidth saving or data rate increase advantages. Better than the non-built-in security OFDM, the non-orthogonally packed sub-carriers in SEFDM bring \ac{ICI}, which complicates signal detections but in turn contributes to secure communications since computationally complex signal detectors would increase the cost of eavesdroppers to detect signals. Previous work in \cite{MOFDM_PIMRC2009} studied the possibility of a similar strategy in physical layer security. The main idea is to overlap two orthogonal OFDM signals. In this case, interference is introduced between two overlapped OFDM signals and eavesdroppers cannot intercept signals without advanced signal detectors. This might be true when computational complexity is the primary concern. However, with the advancement in hardware, brute-force but optimal performance achievable detectors become realistic in consumer-level hardware. Thus, the traditional waveform encryption in \cite{MOFDM_PIMRC2009} is easily broken down and a solution, which can efficiently combat with time-variant multipath fading, multiple user access, deep learning adversarial attack, brute-force offline interception and beamforming leakage for low-cost hardware working in a wide communication range, is in urgent need. 

This work will investigate two waveform dependent defence methods. Firstly, a waveform scaling strategy aiming to increase the number of non-orthogonally packed sub-carriers, can significantly increase the computational complexity of signal detections but in turn prevent eavesdropping and enhance information confidentiality. Secondly, a waveform tuning strategy, related to a waveform bandwidth compression factor adjustment, is proposed to confuse eavesdroppers by misidentifying signals. Deep learning has seen great success in various applications and is believed to be a potential approach to assist eavesdropping. Therefore, a deep learning based eavesdropping attack model is trained to evaluate the robustness of the proposed defence waveforms. Results indicate that by intentionally tuning waveform parameters (i.e. bandwidth compression factor), signal features cannot be correctly identified by eavesdroppers, which results in subsequent eavesdropping detection failures.

\section{Defence Strategy}

\subsection{Defence Waveform}

The proposed defence waveform has self-created \ac{ICI}, which is the essential mechanism of preventing eavesdroppers to accurately identify or detect signals. The principle of the waveform is expressed as
\begin{equation}
X_k=\frac{1}{\sqrt{N}}\sum_{n=0}^{N-1}s_{n}\exp\left(\frac{j2{\pi}nk\alpha}N\right),\label{eq:SEFDM_discrete_signal}\end{equation}
where $s_{n}$ indicates the $n^{th}$ single-carrier symbol within one SEFDM symbol, $N$ is the number of sub-carriers, $k$ denotes time sample index and $\alpha=\Delta{f}\cdotp{T}$ is the bandwidth compression factor where $T$ is the time period of one SEFDM symbol and $\Delta{f}\leq{1/{T}}$ is the sub-carrier spacing. The power of one SEFDM symbol is computed in the following
\begin{equation}\label{eq:SEFDM_square_signal}
\begin{split}
|X_k|^2&=\frac{1}{N}\sum_{n=0}^{N-1}\sum_{m=0}^{N-1}s_{n}s^{*}_{m}\exp\left(\frac{j2{\pi}(n-m)k\alpha}N\right)\\
&=\frac{1}{N}\sum_{n=0}^{N-1}|s_{n}|^2+\\
&\frac{1}{N}\sum_{n=0}^{N-1}\sum_{m\neq{n},m=0}^{N-1}s_{n}s^{*}_{m}\exp\left(\frac{j2{\pi}(n-m)k\alpha}N\right).
\end{split}
\end{equation}

\begin{figure}[ht]
\begin{center}
\includegraphics[scale=0.43]{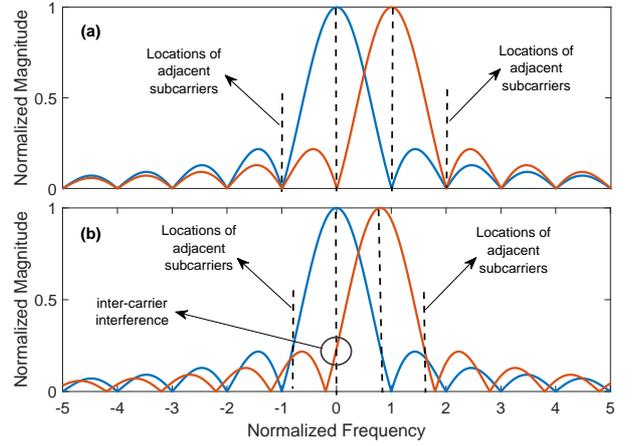}
\end{center}
\caption{Illustration of self-created inter carrier interference within SEFDM signals. (a) OFDM sub-carrier packing. (b) SEFDM sub-carrier packing.}
\label{Fig:AI_encryption_subcarrier_packing}
\end{figure}

The self-created \ac{ICI} within the SEFDM waveform complicates signal detections and therefore increases the cost of eavesdropping. To separate the constructive signal from its self-created destructive interference, variables $m$ and $n$ are introduced in \eqref{eq:SEFDM_square_signal}. The signal part is defined when $m=n$ while the interference part is the term when $m\neq{n}$. It should be noted that the value of $\alpha$ determines the interference term, which is zero when $\alpha={1}$ (i.e. OFDM) while non-zero when $\alpha\neq{1}$ (i.e. SEFDM). An illustration of the non-orthogonal sub-carrier overlapping interference is shown in Fig. \ref{Fig:AI_encryption_subcarrier_packing}, where it clearly shows the \ac{ICI} at each sub-carrier location in SEFDM signals. 

The generation of SEFDM signals is simply performed via IFFT. To remove the parameter $\alpha$ in \eqref{eq:SEFDM_discrete_signal}, a new parameter $M=N/\alpha$ is defined. By padding $M-N$ zeros at the end of each input vector (i.e. a vector consists of $N$ single-carrier symbols), a new vector of input symbols is obtained as
\begin{equation}
s^{'}_{i} = \left\{
  \begin{array}{l l}
    s_{i} & \quad \text{$0{\leq}i<N$}\\
    0 & \quad \text{$N{\leq}i<M$}
  \end{array} \right.
\label{eq:single_IFFT},\end{equation} 
where the value of $N/\alpha$ has to be an integer and simultaneously a power of two, $N/\alpha\in{2^{(\mathbb{N}_{>0})}}$, which allows the IDFT to be implemented by the computationally efficient radix-2 IFFT. The SEFDM signal in a new format is defined as
\begin{equation}
X^{'}_k=\frac{1}{\sqrt{M}}\sum_{n=0}^{M-1}s^{'}_{n}\exp\left(\frac{j2{\pi}nk}M\right),\label{eq:FDM_discrete_signal_single_IFFT}\end{equation}
where $n,k=[0,1,...,M-1]$. The output is cut with only $N$ samples reserved and the rest $M-N$ samples are discarded. Due to the discard of the last $M-N$ samples, ICI is therefore introduced and is treated as a new enhancement solution for physical layer security.

\subsection{Waveform Scaling Defence}

This section will firstly evaluate the defence methodology proposed in \cite{MOFDM_PIMRC2009}, which expects significant performance degradation without using a complex signal detector. An eavesdropper therefore would not extract confidential information from the non-orthogonal signals. The detection of traditional OFDM signals depends on the \ac{MF}, which is essentially an FFT operation at the receiver. The complexity of FFT is acceptable in widely used communication systems, which requires $(N/2)log2(N)$ multiplications and $Nlog2(N)$ additions. For the proposed non-orthogonal signal, the detection relies on the brute-force \ac{ML} detector, which has exponentially increased computational complexity.

In practice, a performance maintained but simpler \ac{SD} detector is used instead of \ac{ML} due to the reduced signal processing complexity by searching for a partial number of solutions. In this case, \ac{SD} is faster than \ac{ML}. However, the complexity of \ac{SD} is random since the search for an optimal solution is related to noise power. Therefore, to get a fair and convincing comparison, the upper bound complexity is considered leading to the search for all possible solutions, which is the case when noise power dominates. In this case, the complexity is fixed and is only related to the number of sub-carriers. This section computes complexity in real-valued operations and only considers the complexity for one OFDM/SEFDM symbol. The computations of multiplication and addition operations are mathematically defined as
\begin{eqnarray}C_{_{SD}}=(\underbrace{\sum_{n=1}^{2N}2^n[2n+1]}_{multiplication})+(\underbrace{\sum_{n=1}^{2N}2^n[2n-1]}_{addition}).\label{eq:SD_mult_add}\end{eqnarray}

\begin{figure}[ht]
\begin{center}
\includegraphics[scale=0.47]{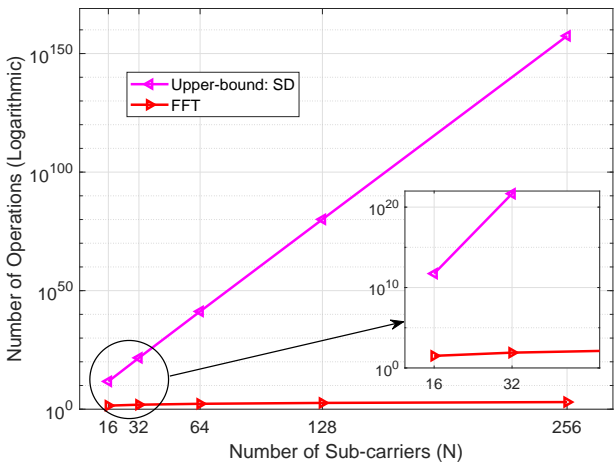}
\end{center}
\caption{The upper bound number (logarithmic) of multiplication operations versus the number of sub-carriers for SEFDM detector (i.e. SD) and OFDM detector (i.e. FFT).}
\label{Fig:AI_complexity_SD_vs_FFT}
\end{figure}

With the breakthrough of low-cost hardware, a complex but powerful detector is no longer a barrier for eavesdroppers to intercept small size signals such as a signal with $N$=12 sub-carriers, which is the size of one resource block in 5G-NR \cite{Erik_book_5G}. Therefore, a straightforward solution to prevent the interception is to make the signal detection harder by scaling up the size of the non-orthogonal signal. The complexity of \ac{SD} is random but it is proportional to the number of sub-carriers. A higher number of sub-carriers can enhance signal encryption by complicating signal detections. Numerical comparisons are presented in Fig. \ref{Fig:AI_complexity_SD_vs_FFT} where only multiplication is considered since its complexity is more concerned in practical systems. For the purpose of illustrations, the number of operations in Fig. \ref{Fig:AI_complexity_SD_vs_FFT} is expressed by a logarithmic scale. Therefore, it is clearly shown that the FFT operation maintains at a low complexity level while the SD complexity increases exponentially. A signal with $N$=256 sub-carriers has an upper bound complexity of $2^{256}$ as shown in Fig. \ref{Fig:AI_complexity_SD_vs_FFT}. Such a large number of mathematical operations would take a significant processing time for the SD detector, which is unrealistic in consumer-level hardware. Thus, the waveform scaling will increase the cost of eavesdroppers to intercept the signals and therefore ensures information confidentiality. 

\begin{figure}[ht]
\begin{center}
\includegraphics[scale=0.47]{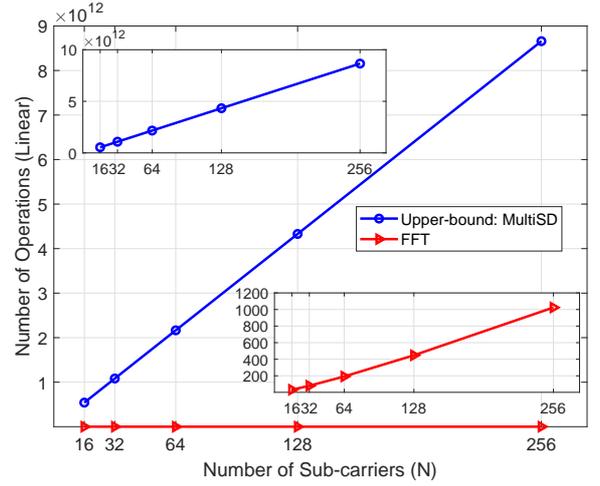}
\end{center}
\caption{The upper bound number (linear) of multiplication operations versus the number of sub-carriers for SEFDM detector (i.e. MultiSD) and OFDM detector (i.e. FFT).}
\label{Fig:AI_complexity_MultiSD_vs_FFT}
\end{figure}

\begin{figure}[ht]
\begin{center}
\includegraphics[scale=0.47]{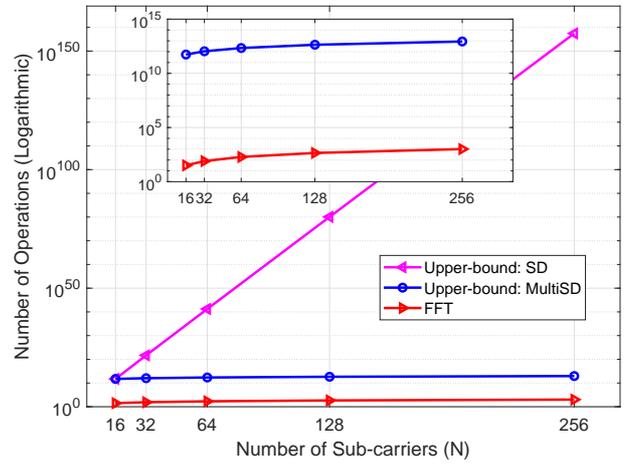}
\end{center}
\caption{The upper bound number (logarithmic) of multiplication operations versus the number of sub-carriers for SEFDM detectors (i.e. SD and MultiSD) and OFDM detector (i.e. FFT).}
\label{Fig:AI_complexity_SD_vs_FFT_vs_MultiSD}
\end{figure}

Waveform scaling is an efficient encryption method to prevent eavesdropping but it also prevents communications between legitimate users. To deal with the detection of such a large size signal, a specially crafted MultiSD detector was proposed in \cite{Tongyang_wincom2017}, which can recover large size non-orthogonal signals with linear computational complexity as shown in Fig. \ref{Fig:AI_complexity_MultiSD_vs_FFT}. The newly designed detector still has higher computational complexity than FFT. However, its multiple-SD architecture enables parallel processing, which is applicable in consumer-level hardware. Its complexity is mathematically expressed as 
\begin{eqnarray}C_{_{M-SD}}=\frac{N}{N_B}(\underbrace{\sum_{n=1}^{2N_B}2^n[2n+1]}_{multiplication})+\frac{N}{N_B}(\underbrace{\sum_{n=1}^{2N_B}2^n[2n-1]}_{addition}).\label{eq:MultiSD_mult_add}\end{eqnarray}

In Fig. \ref{Fig:AI_complexity_SD_vs_FFT_vs_MultiSD}, it clearly shows that the complexity of MultiSD is significantly reduced relative to the traditional SD detector considering the same signal scale. This discovery however endangers the waveform scaling defence since eavesdroppers can intercept signals using the MultiSD detector as well. Therefore, a more clever and robust defence method is needed to cope with the eavesdropping signal detection.

\subsection{Waveform Tuning Defence}

In practice, an eavesdropper has to learn a signal classifier, which can identify different signal formats before any intentional attacks. Existing defence actions for such \ac{AI} dependent eavesdropping would falsify data or labels to prevent accurate classifier training. Without accurate signal identifications, eavesdroppers cannot effectively carry out subsequent attacks. However, these traditional defence mechanisms rely on additional transmissions of fake data and labels, which reduces data throughput between legitimate users. An efficient approach to address the potential detection attack is to design a waveform tuning defence method, which can mislead eavesdroppers into misclassifying the format of signals. The wrong classification of signal formats would result in subsequent detection errors. This solution is to prevent the potential interception from eavesdroppers when the MultiSD detector is known.

\begin{figure}[ht]
\begin{center}
\includegraphics[scale=0.37]{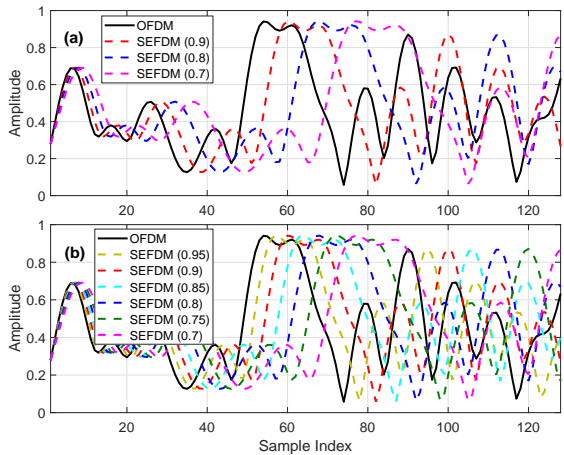}
\end{center}
\caption{Signal feature diversity and similarity visualization by modulating the same QPSK data. (a) Type-I signals. (b) Type-II signals. Values in the bracket indicate the bandwidth compression factor $\alpha$.}
\label{Fig:CNN_feature_combined_G_I_G_II}
\end{figure}

The principle of waveform tuning defence is shown in Fig. \ref{Fig:CNN_feature_combined_G_I_G_II}. It is clearly seen that by tuning the bandwidth compression factors, signal waveforms would have trade-off between diversity and similarity. Type-I shows apparent signal diversity since adjacent signals have evident differences while Type-II shows increased signal similarity because adjacent signals have close features. We would expect that the second type of signals are more difficult to separate from each other than the first type of signals. The same QPSK data is modulated on all the waveforms in Fig. \ref{Fig:CNN_feature_combined_G_I_G_II} merely for signal feature diversity and similarity visualization. For realistic training and testing in the following sections, we would use random QPSK data for each signal waveform. This work assumes that an eavesdropper would automatically learn the features of signals. Therefore, manual feature extractions are not taken into account in this work.

\section{Eavesdropping Model}

\begin{figure}[ht]
\begin{center}
\includegraphics[scale=0.29]{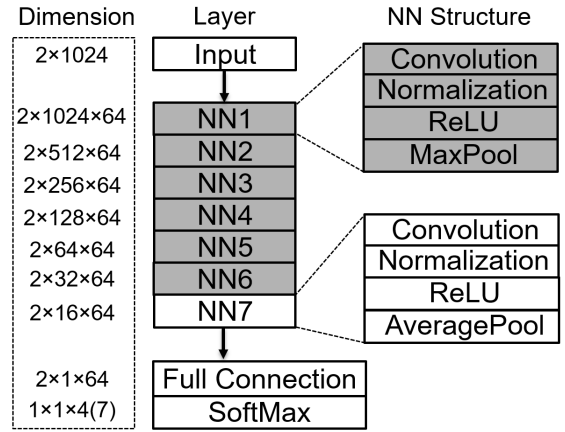}
\end{center}
\caption{CNN classifier neural network layer architecture. }
\label{Fig:CNN_architecture_classification}
\end{figure}

It is assumed that an eavesdropper can train an \ac{AI} signal classifier, which will be used for automatic signal format identification. There is no standardized training methodology for signal classification. Therefore, we apply the deep learning \ac{CNN} model for the eavesdropping signal classifier. The \ac{CNN} architecture \cite{tongyang_VTC2020_DL_classification} is illustrated in Fig. \ref{Fig:CNN_architecture_classification} where seven \ac{NN} layers are designed for signal feature extraction. The first six \ac{NN} layers have the same structure while the last \ac{NN} layer employs Average Pool instead of Max Pool. For signal classification, the CNN model uses a full connection and a SoftMax activation function.

\begin{table}[ht]
\caption{Signal and channel/hardware specifications}
\centering
\begin{tabular}{ll}
\hline \hline
$\mathbf{Parameter}$ & $\mathbf{Specification}$  \\[0.5pt] \hline 
Sampling frequency (kHz) & 200 \\ 
IFFT sample length & 2048 \\ 
Oversampling factor & 8  \\
No. of data sub-carriers & 256 \\ 
Bandwidth compression factor $\alpha$ & 1,0.95,0.9,0.85,0.8,0.75,0.7\\ 
Modulation scheme & QPSK  \\ 
RF center frequency (MHz) & 900  \\ 
Path delay (s) &  [0 9e-6 1.7e-5] \\
Path relative power (dB) &  [0 -2 -10] \\
Maximum Doppler frequency (Hz) & 4 \\
K-factor &  4 \\ 
Frequency offset (PPM) & 2 \\
Omni-directional antenna gain (dBi) & 2 \\\hline \hline
\label{tab:table_signal_channel_specifications}
\end{tabular}
\end{table}

Unlike the single-band signal generation in \cite{tongyang_VTC2020_DL_classification}, this work applies the multi-band signal architecture \cite{Tongyang_wincom2017}, which can confuse eavesdropping signal identification while make legitimate user signal detection possible. The signal for each class (i.e. each $\alpha$) is generated according to Table \ref{tab:table_signal_channel_specifications}. Since over-the-air signals would experience a variety of wireless environments, therefore the signal dataset can be enlarged similar to \cite{tongyang_VTC2020_DL_classification} via data augmentation passing through the time-variant channel models in Table \ref{tab:table_signal_channel_specifications}. Training is operated offline in a computer equipped with an Intel(R) Xeon(R) Silver 4114 CPU (2 processors). Two eavesdropping classifiers, CNN-1 and CNN-2, are trained using the Type-I and Type-II data respectively, which are both distorted by the channel/hardware impairments at a fixed Es/N0=20 dB. A number of 2000 frames (i.e. OFDM/SEFDM symbols) per signal class are generated after the channel/hardware data augmentation. Therefore, there are overall 8000 training frames for the CNN-1. The amount would increase to 14000 training frames for the CNN-2. 

\section{Defence Impact}

\begin{figure}[ht]
\begin{center}
\includegraphics[scale=0.56]{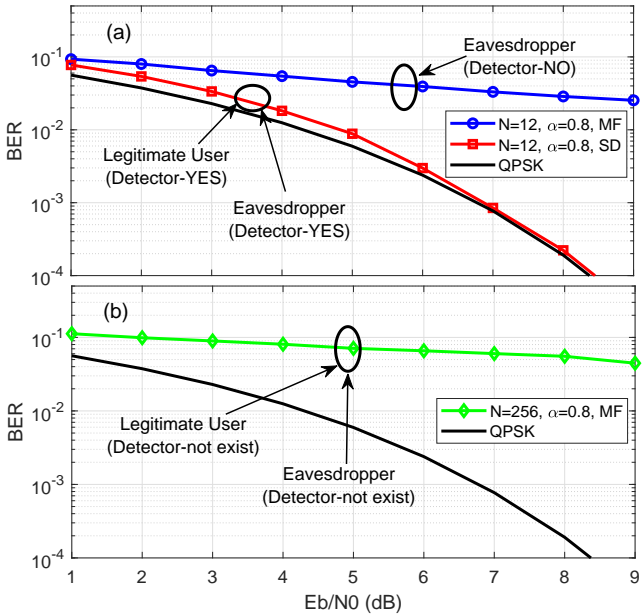}
\end{center}
\caption{Defence impact of waveform scaling.}
\label{Fig:BER_waveform_scaling}
\end{figure}

The original waveform encryption proposal \cite{MOFDM_PIMRC2009} is firstly evaluated in Fig. \ref{Fig:BER_waveform_scaling}(a), in which it assumes that the optimal but complex SD detector is technically challenging for eavesdroppers. Therefore, only simple detectors such as \ac{MF} is applicable. It is clearly seen that the non-orthogonal signal, modulated by 12 sub-carriers, is perfectly recovered by legitimate users using the \ac{SD} detector while it is undetectable by an eavesdropper using \ac{MF}. However, the risk of knowing and applying SD detection for eavesdropping still exists since the rapid advances of hardware making SD detection possible in consumer-level hardware. A straightforward solution is to make detection harder by enlarging the signal size. The performance in Fig. \ref{Fig:BER_waveform_scaling}(b) shows the waveform scaling defence impact on a signal of $N$=256 sub-carriers. The detection of such a signal using SD is impossible since the computational complexity increases exponentially to the upper bound complexity of $2^{256}$. Therefore, it can efficiently prevent eavesdropping. However, it will prevent communications between legitimate users since SD for such a large size signal is not possible for them either. 

\begin{figure}[ht]
\begin{center}
\includegraphics[scale=0.56]{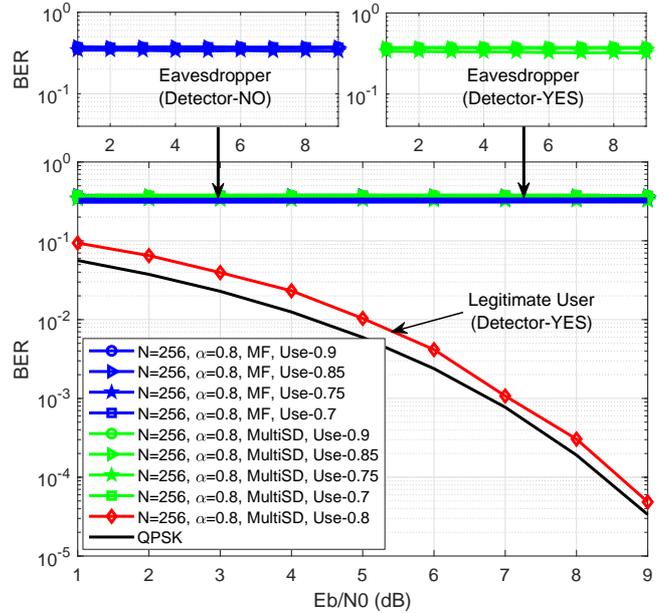}
\end{center}
\caption{Defence impact of waveform tuning.}
\label{Fig:BER_waveform_tuning}
\end{figure}

The proposed waveform tuning can simultaneously deal with eavesdropping encryption and legitimate user signal recovery. Its performance is shown in Fig. \ref{Fig:BER_waveform_tuning}. The target signal waveform is defined by $\alpha$=0.8 while eavesdroppers have no knowledge of signal formats in advance. It clearly shows that the use of incorrect signal detectors (e.g. $\alpha$=0.9, 0.85, 0.75, 0.7) results in great BER performance degradation. It should be noted that the error floors exist for eavesdroppers with or without knowing the MultiSD detector. Only the target legitimate user who knows exact signal formats is able to apply the correct detector (i.e. $\alpha$=0.8) to recover the signal. Thus, the waveform tuning method fundamentally prevents unauthorized interception even the MultiSD detector is leaked to eavesdroppers.

\begin{figure}[ht]
\begin{center}
\includegraphics[scale=0.6]{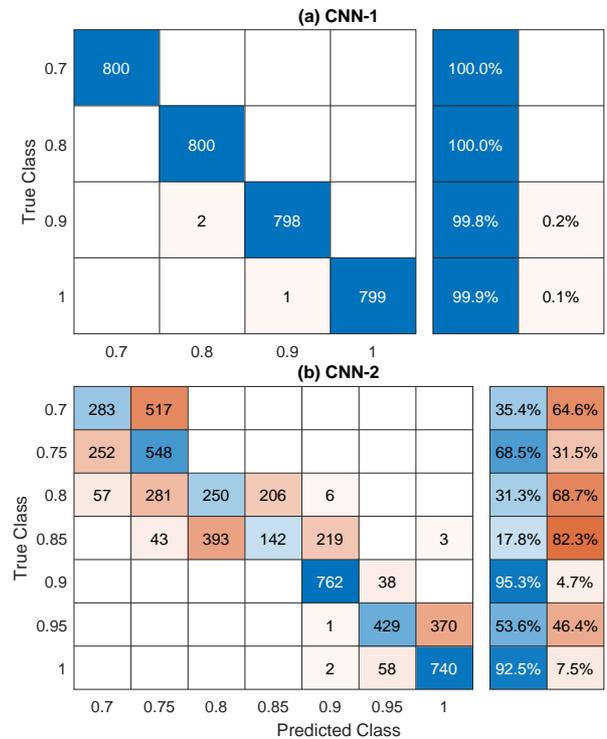}
\end{center}
\caption{Confusion matrix visualization for (a) Type-I and (b) Type-II signals at Es/N0=20 dB. }
\label{Fig:Type_1_2_confusion}
\end{figure}

\begin{figure}[ht]
\begin{center}
\includegraphics[scale=0.42]{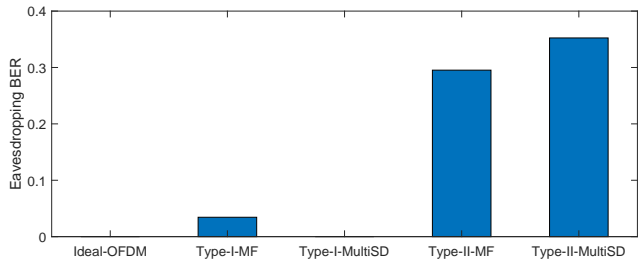}
\end{center}
\caption{Defence impact on eavesdropping BER performance for the target waveform $\alpha$=0.8 at Es/N0=20 dB. }
\label{Fig:BER_defence_performance}
\end{figure}

A realistic waveform tuning impact is shown in Fig. \ref{Fig:Type_1_2_confusion} where confusion matrices, in a similar representation to that of \cite{OShea_classification_2018}, are illustrated for signal classification accuracy. In each sub-figure, classes indicate the bandwidth compression factors $\alpha$, vertical labels indicate true transmitted signal classes and horizontal labels indicate predicted signal classes. Perfect signal classification would show only diagonal elements in each confusion matrix. Therefore, it is visually concluded that Type-I signals yield higher classification accuracy than Type-II signals. The Type-I signal, with less feature similarity, is nearly 100\% accuracy identified by the CNN-1 classifier. By tuning the waveform parameter $\alpha$ to enhance feature similarity, only 56.3\% of Type-II signals are classified into correct signal class.

The misclassification of signal formats, as revealed in Fig. \ref{Fig:Type_1_2_confusion}, would result in significant eavesdropping BER performance degradation as shown in Fig. \ref{Fig:BER_defence_performance}. The BER is zero for traditional OFDM and the MultiSD detected Type-I signal. The Type-I signal has minor BER degradation when \ac{MF} is used. However, the performance of Type-II signal greatly deteriorates whether or not the MultiSD is known. This effectively proves the robustness of the non-orthogonal waveform and its tailored waveform defence strategies in secure communications.

\balance

\section{Conclusion}

This paper investigates the capability of using non-orthogonal waveforms to defend against eavesdropping. Existing proposals on non-orthogonal waveforms rely on the assumption that an eavesdropper cannot intercept signals without complex detectors. However, with the advancement of hardware, the complexity of brute-force signal detection is no longer a barrier. Therefore, a waveform scaling defence strategy is proposed to intentionally further complicate signal detections by scaling up the signal size. The processing complexity increases exponentially and would go up to a level of $2^{256}$. The interception for such signals is impossible to eavesdroppers. However, it also prevents communications between legitimate users since data recovery is also challenging for them. A performance-complexity optimized detector is crafted to deal with large scale non-orthogonal signal detections. However, this endangers secure communications since eavesdroppers can get access to the advanced detector as well. Therefore, a waveform tuning defence strategy is proposed to cope with the aforementioned issue by intentionally tuning waveform parameters. In this case, signals would be tuned to have high feature similarity and eavesdroppers cannot easily identify them. Confusion matrices show that the classification accuracy for diversity dominant signals can approach 100\% while it reduces to 56.3\% when similarity dominates. The low classification accuracy would cause the failure of subsequent signal detections. BER performance reveals the robustness of the waveform tuning strategy, where the misclassification of signals results in detection error floors when the advanced detector is either known or not.

\bibliographystyle{IEEEtran}
\bibliography{Tongyang_Ref}

\end{document}